\documentclass[prl,twocolumn,showpacs,preprintnumbers,amsmath,amssymb]{revtex4}

\usepackage{amscd,graphicx,color,colordvi,bm,nicefrac,psfrag}

\newcommand{\vect}[1]{\mathbf{#1}}
\newcommand{\hvect}[1]{\hat{\mathbf{#1}}}
\newcommand{\vectsym}[1]{\bm{#1}}
\renewcommand{\tensor}[1]{\bm{#1}}

\newcommand{\betaP}{\beta^{\rm P}}
\newcommand{\betaE}{\beta^{\rm E}}
\newcommand{\betaPtensor}{\tensor{\beta}^{\rm P}}
\newcommand{\E}{\mathcal{E}}
\newcommand{\F}{\mathcal{F}}

\begin{document} 


\title{%
  Shocks and slip systems: predictions from a theory of continuum
  dislocation dynamics
} 

\author{Surachate Limkumnerd}
\email{sl129@cornell.edu}

\author{James P. Sethna}
\homepage{http://www.lassp.cornell.edu/sethna/sethna.html}
\affiliation{%
Laboratory of Atomic and Solid State Physics, Clark Hall,\\
Cornell University, Ithaca, NY 14853-2501, USA
}

\date{\today}



\begin{abstract}
Using a recently developed continuum theory of dislocation dynamics,
we derive three new predictions about plasticity and grain boundary
formation in crystals. (1)~There will be a residual stress jump across
grain boundaries and plasticity-induced cell walls, which self-consistently
acts to form the boundary by attracting neighboring dislocations;
we derive the asymptotic late-time dynamics of the grain-boundary
formation process. (2)~At grain boundaries formed at high
temperatures, there will be a cusp in the elastic energy density. 
(3)~In early stages of plasticity, when 
only one type of dislocation is active (single-slip stage~I plasticity),
cell walls will not form; instead we predict
the formation of jump singularities in the dislocation density.
\end{abstract}


\maketitle

Dislocation motion governs the plastic deformation of crystals, as well
as the formation and evolution of grain boundaries in polycrystals.
Acharya and Roy~\cite{RoyAcha05} recently proposed an evolution law
for the density of dislocations in crystals, allowing both glide and climb
(hence appropriate at high temperatures). We subsequently rederived 
this law from the microscopic dynamics and a closure approximation,
proposed a modified law to 
suppress climb, and showed numerically that both laws developed shock
singularities~\cite{LimkSeth06}.
We argued that these shocks are the fundamental origin of the
wall-like structures spontaneously formed by dislocations---providing
a continuum explanation of both
the grain boundaries formed at high temperatures
and the cell walls formed at low temperatures in plastically deformed
crystals.
Wall formation thus emerges naturally from this continuum formulation.
What other properties and predictions naturally emerge from this mesoscale
perspective? Here we discuss three.

(1)~{\em Residual stress.}
The continuum model predicts residual stresses associated with the
formation of grain boundaries and cell walls. It makes detailed
predictions about the asymptotic behavior of the stress and dislocation
density near the walls, and about how the wall's associated stress allows it
to draw in and absorb neighboring dislocations.

(2)~{\em Cusps in energy density.}
The continuum model predicts that the elastic energy associated with the 
residual stress is continuous across a grain boundary (despite the
jumps in the individual stress components). The energy density does have
a cusp singularity.

(3)~{\em Dislocation density jumps in single-slip}.
When appropriately restricted to forbid climb, the model predicts
that materials with only one kind of dislocation will not form cell walls, 
explaining the lack of such structures in systems (discrete-dislocation
simulations and experimental stage~I plasticity) where only one 
slip system is active. Instead, the model predicts that these systems
will form a more subtle jump singularity in the dislocation density.

We will derive these three predictions analytically by mapping the 
dislocation dynamics theory in one dimension onto {\em Burgers equation}. 
We will present numerical simulations to flesh out these predictions.
Finally, we will consider how and whether these predictions are likely to
depend upon the detailed structure of the continuum model, and how physical
mechanisms ignored by the model will likely affect and modify these
striking predictions. We begin by introducing our notation and Acharya
and Roy's derivation of the equations of motion we study.

A complete macroscopic description of the deformation $\vect{u}$ of a
material is given by $\partial_i u_j = \betaE_{ij} + \betaP_{ij}$,
where $\betaE_{ij}$ represents the elastic, reversible distortion and
the plastic distortion tensor $\betaP_{ij}$ describes the irreversible
plastic deformation. The plastic distortion is the result of the net
density of dislocations, described by the Nye dislocation density
tensor~\cite{Nye53,Eshe56,Kose62,Kron58,Mura63}
\begin{equation}
  \rho_{ij}(\vect{x}) = -\varepsilon_{ilm}\partial_l\betaP_{mj} =
  \sum_\alpha t^\alpha_i b^\alpha_j \delta(\vectsym{\xi}^\alpha)
\end{equation}
which measures the net flux of dislocation $\alpha$, tangent to
$\vect{t}$, with Burgers vector $\vect{b}$, in the (coarse-grained)
neighborhood of $\vect{x}$. 

The time evolution of the plastic distortion tensor is given in terms
of a function $J_{ij}= \partial_t \betaP_{ij}$,
where $\varepsilon_{ilm} J_{mj}$ gives the net current for the conserved
Burgers vector of the dislocation density tensor,
$\partial_t \rho_{ij} = -\varepsilon_{ilm}\partial_l J_{mj}$.
The current from a single dislocation moving with velocity $\vect{v}$ is
$J_{ij} = \varepsilon_{iln}t_lb_jv_n\delta(\vectsym{\xi})$, and the
net Peach--Koehler force on a dislocation driving its motion is
$f^{\rm PK}_l = -\varepsilon_{lmn}t_m b_c\sigma_{nc}$
where $\tensor{\sigma}$ is the local stress (due, for example, to the other 
dislocations). In our
derivation of the equation of motion~\cite{LimkSeth06}, we allowed each
dislocation to move independently, and then made a closure approximation
to write $\tensor{J}$ in terms of $\tensor{\rho}$.  Roy and 
Acharya~\cite{RoyAcha05} got the same final result by simply assuming that 
all the dislocations move with the same velocity $\vect{v}$, given by a
constant $(D(\rho)/2)$ times the
force density $\F_l = -\varepsilon_{lmn}\rho_{mc}\sigma_{nc}$ on the 
local dislocations:
\begin{equation}
  \begin{split}
    J^{\rm RA}_{ij} = \frac{D}{2}\,\varepsilon_{ial}
    \F_{l} \rho_{aj} &= -\frac{D}{2}\,
    \varepsilon_{ial} (\varepsilon_{lmn}\rho_{mc}\sigma_{nc})\rho_{aj}
    \\
    &=
    \frac{D}{2}(\sigma_{ic}\rho_{ac} -
    \sigma_{ac}\rho_{ic})\rho_{aj}.
  \end{split}
\end{equation}
A physically natural choice for $D(\rho)$ is proportional to 
an inverse density of dislocation lines (so that the force per 
dislocation drives the motion~\cite{RoyAcha05});
we choose here to discuss the mathematically more convenient choice of
a constant $D$. (Numerically, both give qualitatively similar
evolution~\cite{LimkSeth06}.)
We controlled the microscopic mobility difference between glide and
climb~\cite{LimkSeth06} by
introducing a mesoscale parameter $\lambda$. By setting 
\begin{equation}\label{E:EqnOfMotion}
\partial_t \betaP_{ij} = J_{ij} =
J^{\rm RA}_{ij} - \frac{\lambda}{3}\,\delta_{ij}
J^{\rm RA}_{kk},
\end{equation}
at low
temperatures $\lambda = 1$ removed the trace of $\tensor{J}$ enforcing
volume conservation, and hence forbids climb, while at high
temperatures $\lambda = 0$ allowed for equal mobilities for both glide
and climb.

We now turn to an analysis of the continuum equation in one dimension.
Near a wall singularity (say, perpendicular to $\hvect{z}$)
the dynamics is one-dimensional. The
variations of the stress, plastic strain, and dislocation densities
parallel to the wall asymptotically become unimportant compared to the
variations along $\hvect{z}$ as one approaches the singularity. 
In one dimension the stress $\tensor{\sigma}(\vect{x})$, generally given by a
long-range integral over the neighboring dislocations 
$\tensor{\rho}(\vect{x}')$, can be written as a linear function of
the local plastic distortion $\betaPtensor$:
\begin{equation}\label{E:sigmaOfBeta}
\sigma_{ij} = -\bar{C}_{ijkm} \betaP_{km}
\end{equation}
where 
\begin{equation}\label{E:Cbar}
\bar{C}_{ijkm} = \mu\left(\bar{\delta}_{ik}\bar{\delta}_{jm} +
\bar{\delta}_{im}\bar{\delta}_{jk} +
\frac{2\nu}{1-\nu}\,\bar{\delta}_{ij}\bar{\delta}_{km}\right)
\end{equation}
(not quite equal to the elasticity tensor), and the Kronecker delta
is modified such that $\bar{\delta}_{zz} = 0$.
Hence the evolution law for $\betaPtensor$ simplifies dramatically. The
$\betaP_{zj}$ components do not evolve (except $\betaP_{zz}$ for
$\lambda \ne 0$, which helps enforce volume conservation); the other
components all evolve according to
\begin{equation}\label{E:EOM1D}
  \partial_t \betaP_{ij} = -\frac{1}{2}(\partial_z \E)\,\partial_z\!
  \left(\betaP_{ij} - \frac{\lambda}{3}\,\betaP_{kk} \delta_{ij}\right)
\end{equation}
where we have rescaled the time to set $D = 1$.
The elastic energy density $\E$ in equation~\ref{E:EOM1D} can be
written in terms of the local plastic distortion tensor:
\begin{multline}\label{E:Energy1D}
    \E = \frac{1}{2}\,\bar{C}_{ijkm}\betaP_{ij}\betaP_{km} 
    = \frac{\mu}{2}(\betaP_{xy} + \betaP_{yx})^2 \\ + \mu
    \left({\betaP_{xx}}^2 + {\betaP_{yy}}^2\right) +
    \frac{\mu\nu}{1-\nu}(\betaP_{xx} + \betaP_{yy})^2
\end{multline}

\begin{figure}[htb]
  \centering
  \psfrag{B}{$\betaP_{xz}$}
  \psfrag{E}{$\E$}
  \psfrag{F}{$\F$}
  \psfrag{z}{$z$}
  \includegraphics[height=3.2in]{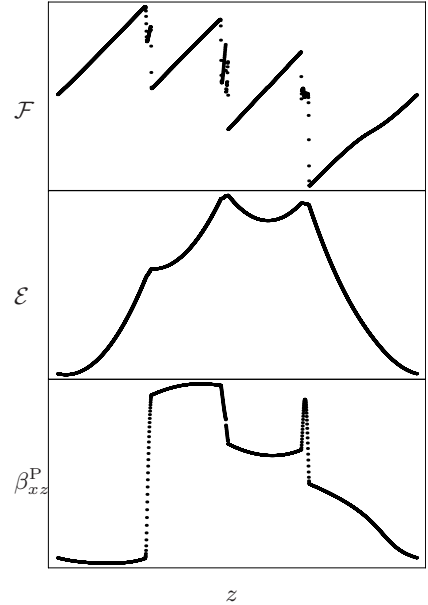}
  \caption{{\bf Cusps and jumps in one dimension.} For the continuum dynamics
  equation allowing both glide and climb, the Peach--Koehler force
  density $\F$ obeys Burgers equation, and hence develops sharp
  jumps (top). The individual components of the plastic distortion
  tensor $\betaPtensor$ (bottom), as well as the stress and strain tensors,
  evolve to also have sharp jumps at the walls; the dislocation
  density hence develops the $\delta$-function singularity associated
  with grain boundary formation. These discontinuities in the residual
  stresses and strains, however, cancel out in the net elastic 
  energy density, which is continuous with only a cusp at the walls (middle).}
  \label{fig:cusps}
\end{figure}

We now specialize to the case of $\lambda = 0$, where glide and climb
are treated on an equal footing (applicable to grain boundary
formation during polygonization at high temperatures, for example). In
this case, equation~\ref{E:EOM1D} tells us that the individual
components of $\betaPtensor$ are all independent of one another,
slaves solely to the evolution of the total stress energy density
$\E$. By contracting equation~\ref{E:EOM1D} with
$\bar{C}_{ijkm}\betaP_{km}$ then using expression~\ref{E:Energy1D},
the time evolution of the strain energy becomes
\begin{equation}\label{E:E_evolution}
  \partial_t\E + \frac{1}{2}\left(\partial_z\E\right)^2 = 0\,.
\end{equation}
Equation~\ref{E:E_evolution} can be cast into the famous Burgers
equation by defining $\F =
\partial_z\E$~\cite{Burgulence,SethRausBouc04,Whit74}:
\begin{equation}
  \partial_t \F + \F\partial_z\F = 0.
\end{equation}
The scalar $\F(z)$ is again the net Peach--Koehler force density on the
local dislocation density $\tensor{\rho}(z)$. Burgers equation is the
archetype of hyperbolic partial differential equations.
Under Burgers equation $\F$ will develop sharp jumps
downward after a finite evolved time, corresponding to cusps in the
energy density $\E$ and leading to jumps in the components of
$\betaPtensor$ (Fig~\ref{fig:cusps}).

At late times, the asymptotic late-time solutions to Burgers equation 
(between the singularities) are linear functions of $z$ whose slopes
decay with time:
\begin{equation}\label{E:Asymp_F}
  \F \sim \frac{z-z_0}{t-t_0}
\end{equation}
The corresponding elastic energy density (continuous across the
singularities) has asymptotic form given by integrating 
equation~\ref{E:Asymp_F}:
\begin{equation}\label{E:Asymp_E}
  \E \sim \frac{1}{2}\frac{(z-z_0)^2}{(t-t_0)} + \E_0
\end{equation}
The individual components of the distortion
tensor (apart from the three time-independent components $\beta_{zj}$)
numerically take the form
\begin{equation}
  \betaP_{ij} \sim \alpha_{ij}\frac{z-z_0}{\sqrt{t-t_0}} + \gamma_{ij}\,,
\end{equation}
which can be shown to be consistent with the evolution law for the 
energy density (equation~\ref{E:Asymp_E}), so long as the coefficients
$\alpha_{ij}$ and $\gamma_{ij}$ obey the relations
\begin{equation}
  \begin{split}
    \bar{C}_{ijkm}\alpha_{ij}\alpha_{km} &= 1\,,\qquad
    \bar{C}_{ijkm}\alpha_{ij}\gamma_{km} = 0\,, \\
    &\bar{C}_{ijkm}\gamma_{ij}\gamma_{km} = 2\E_0\,,
  \end{split}
\end{equation}
for $i,j = x$ or $y$; there are no restrictions on $\alpha_{iz}$
and $\gamma_{iz}$.


There has been extensive work on discrete point dislocation simulations
in two dimensions, where it appears necessary to include more than
one slip system to form walls~\cite{BenzBrecNeed04,BenzBrecNeed05,FourSala96,GomeDeviKubi06,GromBako00,GromPawl93PMA,GromPawl93MSEA,GullHart93}. What does the continuum model
predict for this case? Let
us consider a 2D system (constant along $\hvect{z}$) of straight edge
dislocations with Burgers vector along $\hvect{x}$, described by a single
non-zero component $\rho_{zx}(x,y)$. Such a system has two non-zero
components of the distortion tensor,
$\rho_{zx} = -\partial_x\betaP_{yx} + \partial_y\betaP_{xx}$.

\begin{figure}[htb]
  \centering
  \includegraphics[width=3in]{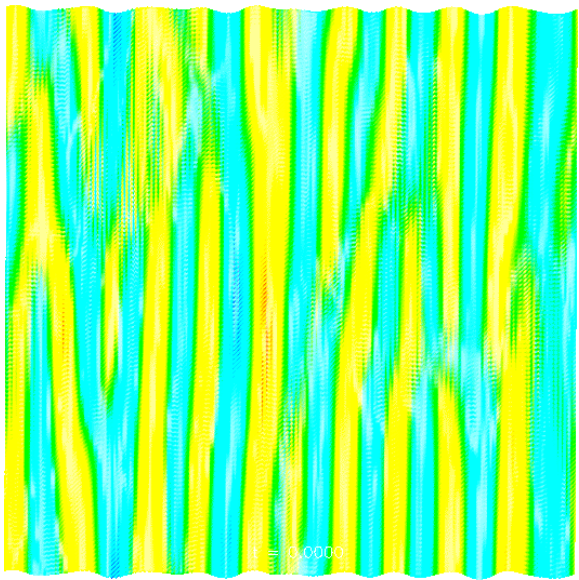}
  \caption{{\bf Continuum of walls.} The dislocation density tensor
  $\rho_{zx}$ evolved allowing both glide and climb from a random
  initial state of edge dislocations along $\vect{t}\sim \hvect{z}$
  with $\vect{b}\sim\hvect{x}$. Notice that the dislocations arrange
  themselves into small-angle tilt boundaries at the lattice scale,
  but do not coarsen; compare to~\cite{BartCarl97}.}
  \label{fig:Supplemental}
\end{figure}

A simulation of this system with Gaussian random initial $\betaP_{yx}$
and $\betaP_{xx}$, allowing both glide and climb, generates a series of
walls of dislocations roughly parallel to the $\hat y$-axis similar to those
seen by Barts and Carlsson~\cite{BartCarl97} in their 2D study of 
single slip with both glide and climb. (See figure~\ref{fig:Supplemental}.)

\begin{figure}[htb]
  \centering
  \includegraphics[width=3in]{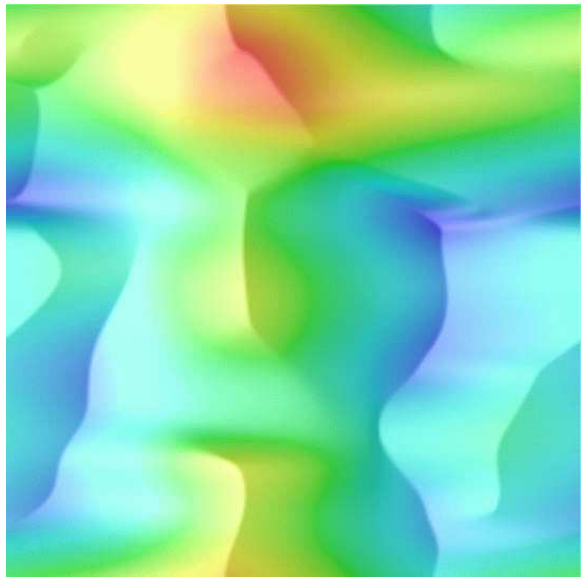}
  \caption{{\bf Cusps formed with one slip system.} Plastic distortion
  tensor $\betaP_{yx}$ formed by climb-free evolution of a Gaussian
  random initial state of edge dislocations pointing along $\hvect{z}$
  with Burgers vector along $\pm \hvect{x}$. Notice that walls do not
  form with one slip system, only cusps in the distortion tensor;
  compare to~\cite{MiguVespZappWeis01a}.}
  \label{fig:rhojump}
\end{figure}

To forbid climb in this case it is convenient
  \footnote{
  Our proposed evolution law equation~\ref{E:EqnOfMotion} suppressed
  climb by removing the trace of the current, hence the term 
  $-\lambda/3\, J_{kk} \delta_{ij}$ in $\partial_t \betaP_{ij}$. This
  choice is inconvenient here, because it introduces new components
  $\betaP_{yy}$ and $\betaP_{zz}$ to the problem, and the corresponding
  dislocation densities $\rho_{zy}$, $\rho_{xz}$, and $\rho_{yz}$. While
  these are allowed by symmetry, they are not part of the discrete
  dislocation simulation.}
to simply choose $\betaP_{xx}\equiv 0$.
Figure~\ref{fig:rhojump} shows the evolution of the distortion field
for climb-free dynamics with a single slip system. In agreement with
experiment and the discrete dislocation
simulations~\cite{BakoGrom99,MiguVespZappWeis01a}, we observe no 
cell wall structures in single slip (which would correspond to jumps
in $\betaPtensor$ in Fig~\ref{fig:rhojump}). Instead, we find a network
of surfaces exhibiting a striking
new singularity: a cusp in the distortion tensor, corresponding to a jump
in the dislocation density.
We can understand this singularity analytically using our mapping
to Burgers equation. With only one non-zero component of $\betaPtensor$,
the energy density is proportional to the square of $\betaP_{yx}$
(eqn~\ref{E:Energy1D}). The energy density satisfies 
eqn~\ref{E:E_evolution} and forms cusps, so the distortion tensor
must also form cusps.



Which of these new predictions of the mesoscale dislocation dynamics model 
can be trusted? Which seem (in retrospect) physically plausible, and which 
are likely artifacts of simplifications made in the modeling process, or
just mathematical curiosities of this particular model? 

(1)~{\em Residual stress.} The primary driving force for dislocation
motion is stress. In cases where dislocations dynamically 
assemble into walls (polygonization at high temperatures, cell wall
formation at low temperatures), it does seem natural in retrospect
to expect that the walls will be associated with stress jumps designed
to attract residual dislocations to the walls. Grain boundaries
formed from the melt when separate growing crystals touch should likely
not be described by the continuum model
\footnote{
One concern we have is that a wall 
with a stress jump can lower its energy by splitting in two; stress
jumps at walls in real materials may be stabilized by a different
mechanism than in the continuum model.}.
These residual stress jumps
must be viewed as a fundamental prediction of the model.

(2)~{\em Cusps in the energy density.} If there are jumps in the stress
at grain boundaries, surely it is natural that there be some singularity in
the energy density. At late stages when the dislocations between grain
boundaries have all been removed, a flat boundary can lower the system energy
by moving into the region of higher energy density. If both glide and
climb are allowed
  \footnote{The glide-only continuum model does have small jumps in the
  energy density.},
and if dislocation mobility is unimpeded (by precipitates, impurities,
lattice pinning, or tangling) this traction will lead the boundaries to move 
until the energy density is continuous across the boundary. Hence,
for mobile walls at high temperatures, it is natural to expect the
energy density to be continuous, and have only cusp singularities.

(3)~{\em Dislocation density jumps in single-slip.} The formation of
dislocation walls in the continuum theory cannot properly be called
a prediction, since wall formation is
well known experimentally. It is, however, a natural consequence of the
hyperbolic form of the equations. The (hitherto unobserved) prediction of
dislocation density jumps in early stages of plasticity when only one
slip system is active is also a natural consequence of hyperbolic 
equations. Our continuum model is guaranteed to lower the net energy
with time, reassuring us that the predicted dislocation density jumps
are energetically favorable and satisfy all compatibility constraints.
They could, however, be smeared by pinning and inhomogeneities
in real systems just as grain boundaries and cell walls are distorted 
by these effects. A smeared dislocation density jump may be more challenging
to identify than a smeared wall of dislocations, perhaps explaining
why these jumps have not yet been seen experimentally.

We thus predict that residual stress is not due to inhomogeneities or other
infelicities in the formation process, but is an intrinsic component of
the formation of grain boundaries and cell walls. We make concrete predictions
about the nature and form of these stresses near grain boundaries, that 
should be testable in colloidal systems or using next-generation X-ray
sources. We provide a mesoscale explanation for one of the
key morphological distinctions between early (stage~I) and later (stage~III)
plasticity. Finally, we predict a new {\em dislocation density jump}
structure in plastically deformed systems with only one active slip system.

\begin{acknowledgements}
We thank Amit Acharya for helpful conversations, and acknowledge 
funding from NSF grants ITR/ASP ACI0085969 and DMR-0218475.
\end{acknowledgements}

\bibliography{ShockSlipSystem}

\end{document}